\documentstyle[12pt,amssymb,epsfig]{article}
\begin{document}
\def \be  {\begin{equation}}
\def \ee  {\end{equation}}
\def \ba  {\begin{eqnarray}}
\def \ea  {\end{eqnarray}}
\def \baa {\begin{eqnarray*}}
\def \eaa {\end{eqnarray*}}
\def \bb  {\begin{thebibliography}}
\def \eb  {\end{thebibliography}}
\setlength{\baselineskip}{0.65cm}
\setlength{\parskip}{0.35cm}
\def    \beq             {\begin{equation}}
\def    \eeq             {\end{equation}}
\def    \beqa            {\begin{eqnarray}}
\def    \eeqa            {\end{eqnarray}}
\def \e {{\rm e}}
\def \as {{\alpha_s}}
\begin{titlepage}
%
\begin{flushright}
BNL-NT-01/20 \\
RBRC-195 \\
August 2001
\end{flushright}

\vspace*{1.5cm}
\begin{center}
\LARGE
{\bf {Next-to-leading order QCD evolution}} 

\medskip
{\bf {of transversity fragmentation functions}}

\vspace*{2cm}
\large 
{Marco Stratmann$^{a}$ and Werner Vogelsang$^{b,c}$}

\vspace*{1.0cm}
\normalsize
{\em $^a$Institut f\"ur Theoretische Physik, Universit\"at Regensburg,\\
D-93040 Regensburg, Germany}\\

\vspace*{0.5cm}
{\em $^b$Physics Department, Brookhaven National Laboratory,\\
Upton, New York 11973, U.S.A.}\\[1mm]

{\em $^c$RIKEN-BNL Research Center, Bldg. 510a, Brookhaven 
National Laboratory, \\
Upton, New York 11973 -- 5000, U.S.A.}
\end{center}

\vspace*{1.5cm}
\begin{abstract}
\noindent
We derive the next-to-leading order splitting kernels
for the scale evolution of fragmentation functions for transversely
polarized quarks into transversely polarized hadrons.  
\end{abstract}
\end{titlepage}
\newpage
%
%
\noindent
Very little is known so far about spin effects in fragmentation
of partons into hadrons. An exception is semi-inclusive production
of $\Lambda$ baryons in $e^+e^-$, $ep$, or $pp$ scattering, where
the polarization of the $\Lambda$ has been 
observed~[1-4]
and been related to
corresponding spin-dependent QCD parton-to-$\Lambda$ fragmentation 
functions~[5-7].

Polarization effects in fragmentation are by themselves interesting
since their study opens up a new perspective on non-perturbative QCD
phenomena in hadron formation. In addition it has been realized that,
if known with sufficient accuracy, spin-dependent fragmentation 
functions can be used as ``polarimeters'' for nucleon 
structure~[8-11].
For instance, if the polarized fragmentation functions for a
transversely polarized quark producing a transversely polarized
$\Lambda$ have been extracted accurately in $e^+e^-$ annihilation,
one can determine with their help the much 
coveted {\em nucleon's transversity
densities} by studying two spin asymmetries with transverse
polarization in $\Lambda$ production in $ep$ or $pp$ collisions.
Behind such a reasoning is of course the QCD factorization 
theorem which states that, for a given produced hadron, 
the fragmentation functions appearing in these scatterings 
are universal, provided of course the process is amenable at all to 
a description in terms of fragmentation functions. This is the
case in the presence of a hard scale in the reaction under
consideration, such as the virtuality $Q$ of the virtual
boson in $e^+e^-$ annihilation or the transverse momentum 
of the $\Lambda$ in $pp\to \Lambda X$. 

It is expected that future experiments will give very precise 
information on spin-dependent fragmentation 
functions~[12-14].
To analyze
such data, an advanced theoretical framework is required. 
According to the factorization theorem the polarized cross section
for, say, $\Lambda$ production in $pp^{\uparrow}\to 
\Lambda^{\uparrow} X$ (the arrow denoting transverse polarization)
is given by a convolution of the form
\begin{eqnarray} 
\label{eq1}
\frac{E\,d\delta \sigma}{dp^3} &\equiv&
\frac{1}{2} \left[\frac{E\,d\sigma}{dp^3}(\uparrow\uparrow) - 
\frac{E\,d\sigma}{dp^3}(\uparrow\downarrow)
\right] \nonumber \\[2mm]
&=& 
\sum_{a,b,c}\,
\int dx_a 
\int dx_b 
\int dz_c \,\;
f_a (x_a,\mu) \;\delta f_b (x_b,\mu)\; 
\frac{E\,d\delta \hat{\sigma}_{ab}^{c}}{dp^3}
(\hat{s},\hat{t},\hat{u}, \mu)\;
\delta D_c^{\Lambda}(z_c,\mu) \; ,
\end{eqnarray}
where $E,p$ are the energy and momentum of the produced $\Lambda$.
It is again assumed that the reaction is characterized by a hard scale
such as the transverse momentum $p_T$ of the $\Lambda$. We denote
by $\hat{s},\hat{t},\hat{u}$ the Mandelstam variables for the
partonic hard-scattering process $ab\to cX$. In Eq.~(\ref{eq1}),
$f_a$ stands for the unpolarized distribution function of parton $a$,
while the $\delta f_b$ are the transversity densities.
The $\delta D_c^H(z,\mu)$ represent the {\em transversity fragmentation}
functions, defined in analogy with the transversity distribution 
functions as 
\begin{equation}\label{eq2}
\delta D_c^{\Lambda}(z,\mu) \equiv D_{c(\uparrow)}^{\Lambda(
\uparrow)}(z,\mu)\; - \; D_{c(\uparrow)}^{\Lambda(\downarrow)}(z,\mu)\; ,
\end{equation}
where $D_{c(\uparrow)}^{\Lambda( \uparrow)}(z,\mu)$ 
$(D_{c(\downarrow)}^{\Lambda(\uparrow)}(z,\mu))$ denotes the 
probability for the fragmentation of a transversely polarized 
parton $c$ to a $\Lambda$ with aligned (anti-aligned) transverse
spin, carrying the fraction $z$ of the parent parton's momentum. Finally,
the sum in Eq.~(\ref{eq1}) is over all contributing partonic channels $a+b\to
c + X$, with $E\,d\delta \hat{\sigma}_{ab}^{c}/dp^3$ the associated 
partonic cross section, defined in complete analogy with the first line of 
Eq.~(\ref{eq1}), the transverse polarizations now referring to partonic ones: 
\begin{equation} \label{eq3}
\frac{E\,d\delta \hat{\sigma}_{ab}^{c}}{dp^3}
\equiv
\frac{1}{2} \Bigg[\frac{E\,d\delta \hat{\sigma}_{ab}^{c}}{dp^3}  
(\uparrow\uparrow) - \frac{E\,d\delta \hat{\sigma}_{ab}^{c}}{dp^3}  
(\uparrow\downarrow)\Bigg] \; .
\end{equation}
The $E\,d\delta \hat{\sigma}_{ab}^{c}/dp^3$ are perturbatively 
calculable thanks to the hard scale involved. 

The factorized form of Eq.~(\ref{eq1}) implies the introduction of
a scale $\mu\sim{\cal O}(p_T)$, 
the factorization scale, that reflects the certain 
amount of arbitrariness in the separation of short-distance and 
long-distance physics embodied in Eq.~(\ref{eq1}).
Even though the parton densities and fragmentation functions
cannot presently be derived from first principles,
their dependence on $\mu$ is calculable perturbatively 
in terms of the ``DGLAP'' evolution equations~\cite{dglap}, allowing 
to relate their values at one scale to their values at any other $\mu$.
Dependence on $\mu$ also arises in the procedure of renormalizing 
the strong coupling constant. Note that in principle one could distinguish 
between factorization scales for the initial and final states and keep 
also the renormalization scale separate; however, for simplicity 
we keep all scales the same. 

In practice, the precision of the framework of Eq.~(\ref{eq1})
largely depends on the perturbative order to which its 
ingredients are calculated. As already pointed out, there are two places 
where perturbation theory enters: first, the $E\,d\delta 
\hat{\sigma}_{ab}^{c}/dp^3$ have the expansion
\begin{equation} \label{eq4}
d\delta \hat{\sigma}_{ab}^{c} = 
d\delta \hat{\sigma}_{ab}^{c,(0)} + 
\left( \frac{\alpha_s}{\pi} \right)
d\delta \hat{\sigma}_{ab}^{c,(1)} + 
\left( \frac{\alpha_s}{\pi} \right)^2
d\delta \hat{\sigma}_{ab}^{c,(2)} + \ldots \;\,.
\end{equation}
We note that currently partonic cross sections involving transverse
polarization in the final state are only known at lowest order, 
except for the reactions $e^+e^-\to q^{\uparrow}\bar{q}^{\uparrow}X$
and  $eq^{\uparrow}\to eq^{\uparrow}X$
for which first-order corrections have been calculated~\cite{cont}.

Secondly, the kernels governing the $\mu$-evolution of the
parton densities and fragmentation functions also enjoy a
perturbative expansion. In this paper we 
present the first-order corrections to the evolution 
of the transversity fragmentation functions defined in 
Eq.~(\ref{eq2}). This seems timely in view of forthcoming 
new experimental information on the production of transversely 
polarized $\Lambda$'s~[12-14].
In addition, the same evolution kernels~\cite{dboer}
drive the evolution of the so-called interference fragmentation
functions introduced in Ref.~\cite{jtt}.

An important observation concerning the $\delta D_c^{\Lambda}$
is that at leading power in the hard scale there is no gluonic transversity
fragmentation function $\delta D_g^{\Lambda}$ due to angular momentum 
conservation and the helicity-flip nature of the 
$\delta D_c^{\Lambda}$~[18-20].
This feature also implies that there is no mixing with gluons in the 
$\mu$-evolution of the $\delta D_q^{\Lambda}$, $\delta D_{\bar{q}}^{\Lambda}$. 
The evolution equations are then most conveniently written in terms 
of the combinations 
\begin{equation}
\delta D_{q,\pm}^{\Lambda} \equiv \delta D_q^{\Lambda} \pm \delta 
D_{\bar{q}}^{\Lambda} \; ,
\end{equation}
for which they read:
\be \label{evolt}
\frac{d}{d\ln \mu^2}\; \delta D_{q,\pm}^{\Lambda}(z,\mu)\;=\; 
\int_z^1 \; \frac{dy}{y}\; \delta P_{qq,\pm}^{(T)} (y,\alpha_s(\mu))\;
\delta D_{q,\pm}^{\Lambda}\left(\frac{z}{y},\mu \right) \; . 
\ee
Here the superscript ``$(T)$'' stands for ``time-like'' and
indicates that we are dealing with a fragmentation function. 
It is instructive to confront Eq.~(\ref{evolt}) with 
the corresponding evolution equations for the (``space-like'')
transversity {\em distribution functions}~\cite{rs,jaffe,ji92,am} 
in, say, the proton,
$\delta q_{\pm} \equiv \delta q \pm \delta \bar{q}$:
\be \label{evols}
\frac{d}{d\ln \mu^2}\; \delta q_{\pm}(x,\mu)\;=\; 
\int_x^1 \; \frac{dy}{y}\; \delta P_{qq,\pm}^{(S)} (y,\alpha_s(\mu))\;
\delta q_{\pm}\left(\frac{x}{y},\mu \right) \; . 
\ee
The evolution kernels $\delta P_{qq,\pm}^{(U)}$ ($U=T,S$) occurring
in Eqs.~(\ref{evolt}), (\ref{evols}) have the perturbative expansion 
\begin{equation} \label{expan}
\delta P_{qq,\pm}^{(U)} (\xi,\alpha_s)\;=\; 
\left( \frac{\alpha_s}{2\pi} \right) 
\delta P_{qq,\pm}^{(U),(0)} (\xi)\; +\; \left( \frac{\alpha_s}{2\pi} \right)^2 
\delta P_{qq,\pm}^{(U),(1)} (\xi)\; +\ldots \; .
\end{equation}
They are in general not identical, 
but are closely related to each other, as can be inferred from 
studies of fragmentation functions in the unpolarized case or in the 
case of longitudinal polarization~[22-25].
To lowest order, the space-like
and time-like splitting functions actually do agree, and there is
also no distinction between the kernels for the evolution of 
the $+$ and $-$ combinations of densities: 
\begin{equation}
\delta P_{qq,\pm}^{(T),(0)} (\xi)\;\equiv\;\delta P_{qq,\pm}^{(S),(0)} (\xi)
\; =\; C_F \left[ \frac{2\,\xi}{(1-\xi)_+}+
\frac{3}{2} \,\delta (1-\xi) \right] \; ,
\end{equation}
where $C_F=4/3$, and the $+$-prescription is defined in the usual way by
\begin{equation}
\int_0^1 dz f(z) \left[ g(z) \right]_+ 
\equiv \int_0^1 dz \left[ f(z)-f(1) \right] g(z) \:\:\: . 
\end{equation}
The splitting function $\delta P_{qq,\pm}^{(S),(0)}$ was derived
in~\cite{bal,am,bl}. The identity of $\delta P_{qq,\pm}^{(T),(0)}$ and 
$\delta P_{qq,\pm}^{(S),(0)}$ is a manifestation of the so-called 
Gribov-Lipatov relation (GLR) \cite{gl} which connects space-like 
and time-like structure functions within their respective physical 
regions $(\xi<1)$. The space-like and time-like leading order 
splitting functions 
are also directly related by analytic continuation through $x=1$:
\ba \label{ACR}
\delta P_{qq,\pm}^{(T),(0)}(z) &=& -z\; \delta P_{qq,\pm}^{(S),(0)} 
\left(\frac{1}{z}\right) \nonumber \\[2mm]
&\equiv& {\cal AC} \Bigg[ \delta P_{qq,\pm}^{(S),(0)} (x)\Bigg] \; ,
\ea
where $x=1/z$ and $z<1$. Eq.~(\ref{ACR}) represents the analytic 
continuation or Drell-Levy-Yan relation (ACR) \cite{dly}. In the 
second line we have introduced the operation ${\cal AC}$ that analytically 
continues any space-like function to $x \rightarrow 1/z >1$ and correctly 
adjusts the normalization. Note that the endpoint contributions 
$\propto \delta (1-z)$ are not subject to the  ${\cal AC}$ operation;
however, they are necessarily identical in the space-like and time-like cases.

Both the GLR and the ACR are known to be violated beyond the lowest 
order~\cite{cfp,sv,blum}. 
However, the ACR is based on symmetries of tree diagrams under crossing, 
and therefore its breaking at next-to-leading order (NLO) is essentially of
kinematical origin within a given regularization prescription, 
as was shown in Ref.~\cite{sv}. The main issue here is an extra
factor of $z^{-2 \epsilon}$ (in $\overline{\rm{MS}}$-scheme
dimensional regularization with $n=4-2 \epsilon$ space-time
dimensions) in phase space for the time-like case which, when 
combined with terms singular at $\epsilon=0$, generates
extra terms $\propto \ln z$ in the final answer of the time-like 
NLO splitting functions. It is then fairly straightforward to go through
the calculation of Ref.~\cite{tsplit} of the space-like NLO transversity 
splitting functions, and to identify in each contributing Feynman diagram
the ACR breaking effects in the procedure of analytic continuation to $x>1$. 

Combining all extra terms, we obtain in the $\overline{\rm{MS}}$ scheme:
\be \label{ACRNLO2}
\delta P_{qq,\pm}^{(T),(1)} (z) 
\;=\;  {\cal AC} \Bigg[ \delta P_{qq,\pm}^{(S),(1)} (x) \Bigg] \;+\; 
 \beta_0 \;\delta P_{qq,\pm}^{(S),(0)}(z)\ln z \; ,
\ee
(for $z<1$) 
where $\beta_0 = 11/3\, C_A - 4/3 T_R n_f$ with $C_A=3$, $T_R=1/2$,
and the number of active flavors, $n_f$. The last term in 
Eq.~(\ref{ACRNLO2}) obviously represents the breaking of the
ACR. It is worth pointing out that the structure of that term
with its proportionality to both $\beta_0$ and the lowest-order
splitting function is that of a typical factorization scheme 
transformation. In other words, we could choose a (non-$\overline{\rm{MS}}$)
factorization scheme in which the time-like transversity splitting
functions would be given by ${\cal AC} \Bigg[ \delta P_{qq,\pm}^{(S),(1)} 
(x) \Bigg]$, without any extra term, so that no breaking of the 
ACR would occur. This possibility was first demonstrated for the 
unpolarized~\cite{beau} and longitudinally polarized~\cite{sv} cases, 
which are more general
in the sense that singlet mixing is present there. In the following,
we do stay within the more conventional $\overline{\rm{MS}}$ scheme, however.

Inserting the explicit result of~\cite{tsplit} for the NLO
$\delta P_{qq,\pm}^{(S),(1)} (x)$, performing the analytic
continuation, and adding the endpoint contributions,
we arrive at the final result for $\delta P_{qq,\pm}^{(T),(1)} 
(z)$. We first define
\begin{eqnarray} 
\delta p_{qq}^{(0)} (z) &\equiv& \frac{2z}{(1-z)_+} \; , \\
{\cal{S}}_2(z) &=& \int_{\frac{z}{1+z}}^{\frac{1}{1+z}} \frac{dy}{y} 
\ln \left(\frac{1-y}{y}\right) \nonumber \\
&=& -2\, {\rm Li}_2 (-z)-2 \ln z \ln (1+z)+\frac{1}{2} \ln^2 z-
\frac{\pi^2}{6} \; ,
\end{eqnarray}
where ${\rm Li}_2 (z)$ is the dilogarithm function. Then
\begin{equation}
\delta P_{qq,\pm}^{(T),(1)} (z) \;\equiv\; \delta P_{qq}^{(T),(1)} 
(z) \;\pm\; \delta P_{q\bar{q}}^{(T),(1)} (z) \; ,
\end{equation}
where
\begin{eqnarray} 
\delta P_{qq}^{(T),(1)} (z) &=& C_F^2 \Bigg[ 1-z + \left( \frac{3}{2} + 
2 \ln (1-z) -2 \ln z \right) \ln z \;  \delta p_{qq}^{(0)}(z) \nonumber \\[2mm]
&&\hspace*{0.65cm} + \left. \left( \frac{3}{8} -\frac{\pi^2}{2} + 6\zeta (3) 
\right) \delta (1-z) \right] \nonumber \\[2mm]
&+& \frac{1}{2} C_F C_A \left[ - (1-z) + \left( \frac{67}{9} + \frac{11}{3} 
\ln z + \ln^2 z - \frac{\pi^2}{3} \right) \delta p_{qq}^{(0)}(z) \right. 
\nonumber \\[2mm]
&& \hspace*{1.6cm} + \left. \left( \frac{17}{12} + \frac{11 \pi^2}{9} -
6 \zeta (3) \right) \delta (1-z) \right] \nonumber \\[2mm]
&+&\frac{2}{3} C_F T_R n_f  \left[ \left( - \ln z -\frac{5}{3} \right) \delta 
p_{qq}^{(0)}(z) - \left( \frac{1}{4} + \frac{\pi^2}{3} \right) \delta 
(1-z) \right] \; , \label{ppp} \\[4mm]
\delta P_{q\bar{q}}^{(T),(1)} (z) &=& C_F \left( C_F - \frac{1}{2} C_A \right)
\Bigg[ -(1 -z) + 2 {\cal{S}}_2 (z) \delta p_{qq}^{(0)}(-z) \Bigg] \; , \label{ppm}
\end{eqnarray}
where $\zeta (3) \approx 1.202057$. 

We are also in the position now to obtain the difference 
$\delta P_{qq,\pm}^{(T),(1)} (z) - \delta P_{qq,\pm}^{(S),(1)} (z)$ 
which is non-zero if the GLR is violated. Using Eqs.~(\ref{ppp}) and
(\ref{ppm}) along with the results of Ref.~\cite{tsplit}, we find:
\be
\delta P_{qq,\pm}^{(T),(1)} (z) - \delta P_{qq,\pm}^{(S),(1)} (z)\;=\;
C_F^2 \ln z \; \delta p_{qq}^{(0)}(z) \; \Big(3+4 \ln (1-z)-2 \ln z
\Big) \; .
\ee
Note that as in the unpolarized and longitudinally polarized 
cases~\cite{cfp,sv} the violation of the GLR only occurs in the $C_F^2$ 
part of the splitting function.

For numerical applications it is convenient to have the Mellin-$n$ 
moments of the NLO splitting functions $\delta P_{qq,\pm}^{(T),(1)} (z)$,
defined as
\be
\delta \tilde{P}_{qq,\eta}^{(T),(1)}(n) \; \equiv \;
\int_0^1 \, dz \,z^{n-1} \;\delta P_{qq,\pm}^{(T),(1)} (z) \; .
\ee
We obtain:
\begin{eqnarray}
\delta \tilde{P}_{qq,\eta}^{(T),(1)}(n) &=& 
\delta \tilde{P}_{qq,\eta}^{(S),(1)}(n) \; + \;
2 \,C_F^2 \Big[ 4 S_1(n) -3 \Big] \left[ \frac{\pi^2}{6} - S_2(n)
\right] \nonumber \\[2mm]
&=& C_F^2 \left[ \frac{3}{8} + 
\frac{1-\eta}{n (n+1)} + 3 S_2 (n) - 4 S_1 (n) \left( 3 \,S_2 (n) - 
S'_2 \left( \frac{n}{2} \right) \right) - 8 \tilde{S}(n) + S'_3 
\left(\frac{n}{2}\right) \right.  \nonumber \\[2mm]
&& \left.\hspace*{0.65cm} 
+\frac{\pi^2}{3} \Big( 4 S_1(n) -3 \Big) \right]\nonumber \\[2mm]
&+& \frac{1}{2} C_F C_A \left[ \frac{17}{12} - \frac{1-\eta}{n (n+1)}
- \frac{134}{9} S_1 (n) + \frac{22}{3} S_2 (n) \right. \nonumber \\[2mm]
&&\left. \hspace*{1.5cm}
+ 4 S_1 (n) \left( 2 S_2 (n) - S'_2 \left(\frac{n}{2}\right) \right)
+ 8 \tilde{S}(n) - S'_3 \left(\frac{n}{2}\right) \right]\nonumber \\[2mm]
&+& \frac{2}{3} C_F T_R n_f \left[ - \frac{1}{4} + \frac{10}{3} S_1 (n) - 
2 S_2 (n) \right]\; ,
\end{eqnarray} 
where $\eta\equiv\pm$, and where in the first line we have expressed
the result in terms of the moments of the space-like NLO transversity 
splitting functions~\cite{tsplit,new1,new2}. 
The sums appearing above are defined by
\begin{eqnarray}
S_k (n) &\equiv& \sum_{j=1}^n \frac{1}{j^k} \nonumber \; , \\
S'_k \left(\frac{n}{2}\right) 
&\equiv& 2^{k-1} \sum_{j=1}^n \frac{1+(-1)^j}{j^k}
\nonumber \; \\
\tilde{S} (n) &\equiv& \sum_{j=1}^n \frac{(-1)^j}{j^2} S_1 (j) \; .
\end{eqnarray}
Their analytic continuations to arbitrary Mellin-$n$ (which depend on 
$\eta$) can be found in \cite{grv}.

In summary, we have presented the next-to-leading order 
kernels for the evolution of trans\-ver\-sity fragmentation functions.
Our results will become useful in analyses of future precision
data sensitive to leading-twist transverse-spin effects in fragmentation.

\section*{Acknowledgments}
We are grateful to D.\ Boer for pointing out to us the relevance of
this project and for helpful discussions.  
M.S.\ thanks RIKEN and Brookhaven National Laboratory for hospitality
and support.
W.V. is grateful to RIKEN, Brookhaven National Laboratory and the U.S.
Department of Energy (contract number DE-AC02-98CH10886) for
providing the facilities essential for the completion of this work.
\newpage

%
\end{document}